# Laser wakefield acceleration with mid-IR laser pulses

D. Woodbury,[1] L. Feder,[1] V. Shumakova,[2] C. Gollner,[2] R. Schwartz,[1] B. Miao,[1] F. Salehi,[1] A. Korolov,[1] A. Pugžlys,[2] A. Baltuška,[2] and H. M. Milchberg[1,*]

[1] Institute for Research in Electronics and Applied Physics, University of Maryland, College Park, MD 20742, USA
[2] Photonics Institute, Vienna University of Technology, Gusshausstrasse 27-387, A-1040, Vienna, Austria
*Corresponding author: milch@umd.edu

We report on the first results of laser plasma wakefield acceleration driven by ultrashort mid-infrared laser pulses (λ = 3.9 μm, 100 fs, 0.25 TW), which enable near- and above-critical density interactions with moderate-density gas jets. Relativistic electron acceleration up to ∼12 MeV occurs when the jet width exceeds the threshold scale length for relativistic self-focusing. We present scaling trends in the accelerated beam profiles, charge and spectra, which are supported by particle-in-cell simulations and time-resolved images of the interaction. For similarly scaled conditions, we observe significant increases in accelerated charge compared to previous experiments with near-infrared (λ=800 nm) pulses.

Ultrashort mid-infrared laser sources have recently garnered attention in strong-field physics applications, demonstrating high harmonic generation up to keV energies [1], K-alpha x-ray production from solid targets [2], and both nonlinear compression [3] and supercontinuum generation [4] in bulk material with anomalous dispersion. The relativistic nonlinearity in laser-plasma interaction is enhanced by both the increased normalized vector potential, which scales as $a_0 \propto \sqrt{I\lambda^2}$, and from a decrease in the critical plasma density as $n_{cr} \propto \lambda^{-2}$. At long wavelengths, achieving reproducible near-critical density targets with tunable profiles is greatly simplified, since exploding solid targets, aerogels or complex pressure boosted gas jets can be replaced by simpler gas jets pulsed at high repetition rates or run continuously. Near the critical density, laser pulses experience lowered thresholds for relativistic self-focusing, onset of parametric instabilities, and enhanced absorption and coupling to plasma oscillations [5], which in turn enable such schemes as magnetic vortex acceleration of plasma ions [6,7].

In laser wakefield acceleration (LWFA) of electrons [8], relativistic self-focusing has long been used to both increase the laser intensity and promote self-guided propagation for extended interaction lengths [9]. The optical nonlinearity responsible for self-focusing arises from the relativistic mass increase of electrons moving in the intense laser field. Above a critical power threshold, $P_c = 17.4\, n_{cr}/n_e$ (GW) [10], where $n_e$ is the electron density, relativistic self-focusing overcomes diffraction and the pulse can collapse. For self-focusing in uniform plasma that occurs in less than the Rayleigh range $z_R$, the collapse distance is given approximately by the self-focusing scale length $\ell_{sf} = z_R(P/P_c)^{-1/2}$ for phase fronts converging due to nonlinear phase picked up over this length. Collapse is arrested by plasma blowout—where electrons are expelled by the pulse to form a highly nonlinear plasma wake—and subsequent injection of background electrons into the wakefield can accelerate them to relativistic energies. Laser plasma acceleration experiments relying on self-focusing have usually required large, multi-terawatt lasers.

Recently we demonstrated that very high density, cryogenically cooled gas jets enable near-critical density laser-plasma interaction for Ti:Sapphire lasers at λ=0.8μm, lowering the threshold for relativistic self-focusing and allowing sub-terawatt pulses to drive highly nonlinear plasma waves in the self-modulated laser wakefield (SM-LWFA) regime [11,12]. In this Letter we demonstrate, for the first time, laser wakefield acceleration using femtosecond mid-IR laser pulses (λ=3.9μm). We study electron acceleration in gas jet targets from a few percent to over twice the critical density, and perform a detailed scan of the power and length thresholds for acceleration. We also present images of the self-focusing process with a synchronized optical probe at λ=650nm, for which all the plasmas studied are well below critical density.

Experiments were conducted with a hybrid OPA/OPCPA laser system [13], which produces ∼25 mJ, ∼100 fs pulses at λ=3.9 μm at a repetition rate of 20 Hz. The experimental setup is shown in Fig. 1. The laser was focused with an off axis $f/5$ paraboloid to a 30 μm FWHM spot size, as determined by a knife edge scan. Due to losses on routing mirrors, the maximum pulse energy on target was 23 mJ, corresponding to a peak vacuum intensity ∼2x10$^{16}$ W/cm$^2$ and a normalized vector potential of $a_0$~0.5. The pulse energy on target was decreased in steps by inserting absorptive glass plates into the beam. The FWHM pulse duration was in the range 105–130 fs, depending on the inserted plates, as measured with a SHG FROG.

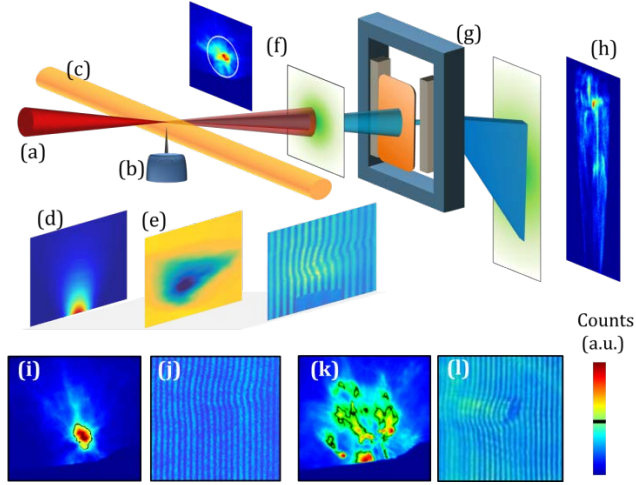

Fig. 1. Top, Schematic of experiment. Mid-IR (3. 9 µm, 105 fs, 10-25 mJ) laser pulses (a), are focused on to the output of a 150 µm orifice gas jet backed at high pressure (b). A synchronized 650 nm probe (c) allowed extraction of both neutral gas densities (d) and plasma induced phase (e). Electron beam profiles (f) were collected by imaging a Lanex screen to a low noise detector. A removable electron spectrometer (g) selected a portion of the beam and dispersed it with a dipole field onto a Lanex screen to capture electron spectra (h). Bottom, example electron beam profiles and interferograms at $P/P_c = 4$ (i, j) and $P/P_c = 10$ (k, l), where the latter images show relativistic multi-filamentation. Beam images are scaled to the maximum counts, and the black line indicates a 50% contour, which for the single beam on the left has a divergence ~200 mrad.

Our hydrogen gas jet target, described in [11], was not cryogenically cooled in the present experiment because for λ=3.9µm the critical density regime ($n_{cr}$~7×$10^{19}$ cm$^{-3}$) was easily achieved with moderate gas densities. By adjusting the backing pressure of the 150 µm orifice diameter nozzles and the distance of the laser axis above the orifice, we achieved tunable peak H$_2$ densities from 1.8x$10^{18}$ to 8x$10^{19}$ cm$^{-3}$, with near-Gaussian density profiles and FWHM widths $d_{FWHM}$ ~ 250 - 1000 µm. When fully ionized, the target density in the laser path spans 0.05$n_{cr}$–2.2$n_{cr}$ at λ=3.9 µm.

Neutral gas and plasma profiles were probed with λ=650 nm, 130 fs pulses from an OPA synchronized with the λ=3.9µm pulses, and imaged by an *f*/2 achromatic lens telescope to a compact Nomarski interferometer [14]. A LANEX scintillating screen, located 7.5 cm beyond the gas jet and shielded from laser exposure by 100 µm thick aluminum foil, was imaged to a low noise CCD camera to capture full electron beam profiles (Fig. 1(f)). The magnetic spectrometer consisted of a 500 µm slit 7.5 cm beyond the jet followed by interchangeable permanent magnets with effective field strengths of 0.065 or 0.013 T, providing spectra in the ranges 750 keV–6 MeV and 2–12 MeV, with the LANEX screen 17.5 cm beyond the jet (Fig 1(g-h)).

Electron acceleration above ~500 keV, the low energy detection limit of our beam profile monitor, is observed over a wide parameter range and is summarized in Figure 2. The unifying theme of this plot is that acceleration occurs for self-focusing lengths less than the target width, $\ell_{sf} < \sim d_{FWHM}$ , with the straightforward interpretation that electron acceleration requires self-focusing of the beam within the axial extent of the plasma in order to drive a plasma wake capable of self-injection. The accelerated electron bunches have a divergence of ~200 mrad for $P/P_c$ ~2, and show increasing divergence as $P/P_c$ increases. For $P/P_c > $ ~10 (for which $n_e > n_{cr}$), the electron beam profile exhibits the signature of relativistic multi-filamentation of the drive pulse (modulational instability), which can also be inferred from interferograms (see Fig. 1(k, l)). Near the $P/P_c$ threshold, beam profiles also exhibit a "halo" of accelerated charge at high divergence (~1 rad).

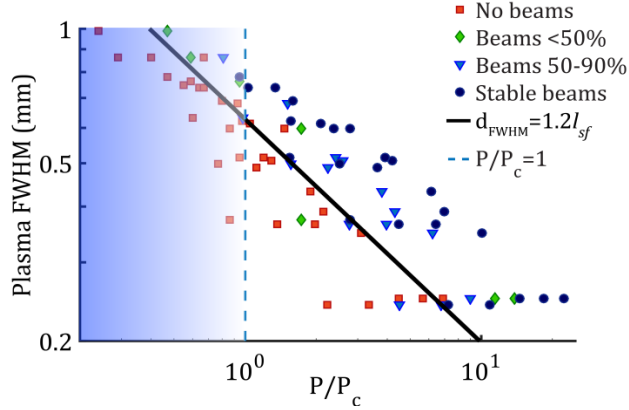

Fig. 2. Plot demonstrating the onset of electron acceleration for a variety of target conditions. Results are grouped by the percentage of 20 shots which produced beams., where beam presence is defined as >2 pC of whole beam charge >0.5 MeV above background noise on an individual shot, The solid line indicates the threshold relationship between the gas jet width $d_{FWHM}$ and the self-focusing length, while the shaded area and dotted line mark the range of $P/P_c$ where appreciable self-focusing is not expected.

    Electron spectra averaged over 20 shots for selected conditions are shown in Figure 3, and display a Maxwellian energy dependence, with effective temperatures ranging from ~0.5 MeV to >2 MeV. Peak energies extend beyond 12 MeV (the resolution limit of the detector). Under all conditions, charge increases with increasing pump energy.
The effect of the plasma density, and hence critical power, is illustrated in Fig. 3(a), which shows accelerated electron spectra as a function of density for a 23 mJ pump incident on a gas target with $d_{FWHM}$ ~700 μm. Electron charge increases quickly as the peak density increases to $0.2n_{cr}$, but then drops as the density is further increased. For the same laser energy, 23 mJ, Figure 3(b) shows total charge per steradian accelerated above 2 MeV vs. plasma density for various jet widths. It is seen that the plasma density giving peak accelerated charge depends on the jet width. Beyond this peak, charge and effective electron temperature both decrease. This suggests that for each jet width, a particular density (and value of $P/P_c$) optimizes the acceleration process. Comparing the optimal (highest charge) cases at each jet width, we find that all share a ratio of jet width to self-focusing length of $d_{FWHM}/\ell_{sf}$~2 (with all accelerated beams lying in ~1.2 < $d_{FWHM}/\ell_{sf}$ < ~3). Near the length onset threshold $d_{FWHM}/\ell_{sf}$~1.2 we also observe highly structured spectra that show quasi-monoenergetic peaks on top of the Maxwellian spectrum for individual shots (e.g. Fig. 1(h)), in line with previous work [11].

    Applying the divergences measured from full beam profiles to the electron spectra, we estimate charge up to ~850 pC above 650 keV for an average of 20 shots, with ~300 pC above 2 MeV. This is consistent with directly integrating the full beam profile and using an estimate of the energy-dependent LANEX response [15, 16], which gives an approximate maximum beam charge ~1 nC at energies >500 keV. Using this full beam estimate, it is again seen that the total accelerated charge peaks for $d_{FWHM}/\ell_{sf}$~2.
A possible explanation for the correlation of charge with self-focusing length, as well as the appearance of quasi-monoenergetic peaks, is shown in Fig. 4 (a). For simplicity, the density profile (dashed curve) is imagined as several stepped regions where a fixed energy pulse is below or above the critical power. The self-focusing length $\ell_{sf}$ defines an effective collapse location in one of these regions, which will move forward as the critical power is increased. For collapse occurring in a region of low density plasma where $P/P_c < 1$, the nonlinearly focused beam quickly diffracts before self-modulation can drive injection into the laser wake, as depicted by the sketch of the expanding beam on the far right. As the collapse moves to areas of higher density (either through an increased jet width or shorter $\ell_{sf}$, a weak nonlinearity can partially guide the focused beam (center beam sketch), leading to self-modulation. At threshold, a limited amount of charge will be injected and accelerated before the beam starts to diffract and injection stops, leading to quasi-monoenergetic peaks of charge accelerated in the laser's linear wakefield. Collapsing earlier in the jet increases the guiding distance (left beam sketch), and when overlapped with the rear density gradient leads to increased charge due to downramp injection [17]. For collapse within the peak of the jet, charge may decrease with the onset of relativistic multi-filamentation and reduced downramp injection, while shorter dephasing lengths [9] may reduce the effective temperature.

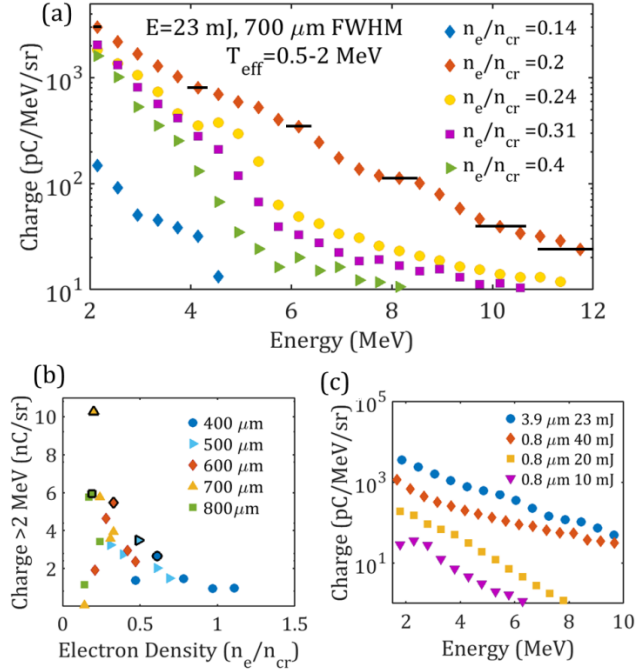

Fig. 3. (a) Electron spectra for jet $d_{FWHM}$ ~700 $\mu m$, 25 mJ pump and varying peak electron densities; bars indicate energy uncertainty, as determined by ray tracing (b) Charge accelerated above 2 MeV by a 25 mJ pump vs. peak density for a variety of target widths. For each target width, the maximum value of charge accelerated is outlined in black. (c) Optimized electron spectra from current $\lambda$=3.9$\mu$m experiment compared to spectra from $\lambda$=800 nm experiment reported in [11], both for $n_e$~$0.25n_c$. All $\lambda$=3.9 $\mu$m spectra are averaged over 20 shots.

The larger plasma structure and lower plasma density than in previous studies at $\lambda = 800\ nm$ [11] enables more detailed transverse interferometric probing of the plasma generated by the wakefield drive pulse, using the $\lambda$=650nm probe and a 2-D phase unwrapping algorithm [18]. By adjusting the delay between probe and drive pulse, we recorded time resolved images of the relativistic self-focusing process for a variety of target conditions. Figure 4(b) shows averaged phase images for two jet widths (300μm and 450μm) and three different interactions ($P/P_c$ ~3, ~7.5 and ~14) at two probe delays: ~500 fs after the drive pulse enters the jet and after the pulse has completely left the jet. Each panel indicates the calculated value of $\ell_{sf}$ for its conditions, and it is seen that the phase images of the interaction follow the same trends.

We note that earlier work with a pressure boosted jet [19] imaged dynamics of a self-focusing pulse in a near-critical interaction, but for $P/P_c$ >200. As a result, the pulse abruptly collapsed on the density up-ramp and deposited its energy into a population of divergent, thermal electrons driving an opaque shock structure. Here, owing to the larger values of $l_{sf}$, we observe details of a more gradual collapse accompanied by electron acceleration to MeV energies.

As seen in Fig. 3(c), the charge in $\lambda$=3.9 μm laser-driven electron beams at optimized conditions significantly exceeds charge in beams driven at $\lambda$=800 nm [11]. When the comparison is for laser pulses of the same energy (20 mJ), the charge increase factor is ~20, and ~100 for the same peak power (200 GW). If we assume that the number of accelerated charges scales as the background density times the volume of a plasma bucket, $N_{acc} \propto n_e \lambda_p^3 \propto n_e^{-1/2}$, then for operating at a fixed fraction of the critical density, the scaling is $N_{acc} \propto \lambda$, which would predict a 5× charge enhancement, lower than observed. In the self-modulated regime, where the laser envelope overlaps multiple buckets, this scaling may be modified. In addition, due to the observed importance of the self-focusing length, it is possible that the jet width in the previous $\lambda$=800 nm experiment was not optimized for maximum charge.

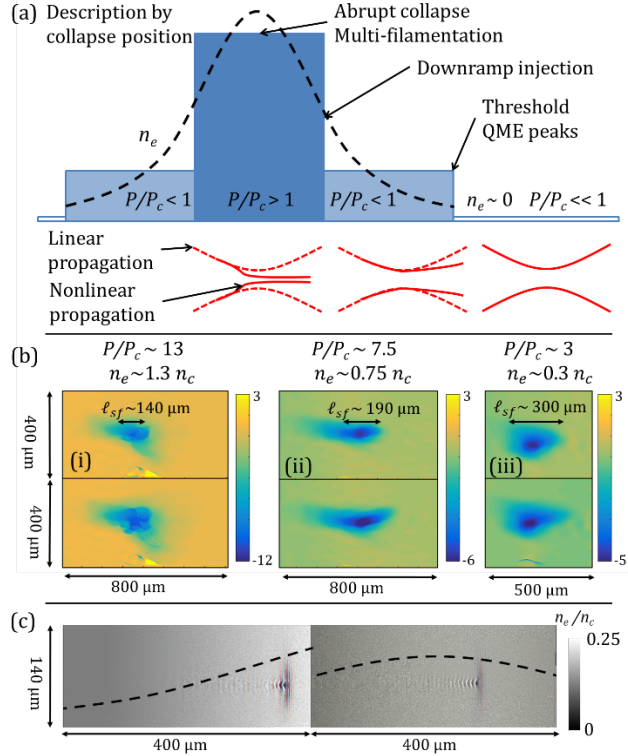

Fig. 4. (a) Qualitative behavior of electron beams for nonlinear collapse at different locations within the plasma density profile. Self-guiding only occurs if the pulse collapses in areas where $P/P_c > 1$, and will diffract elsewhere, though with partial confinement for $P/P_c \lesssim 1$. The guiding length affects electron injection, leading to a scale length threshold and charge increases for focusing on the density downramp. (b) Drive laser-induced plasma phase images (average of 20 shots) during (top) and after (bottom) interaction with jet widths (i-ii) $d_{FWHM}$ = 450 μm or (iii) $d_{FWHM}$ =300 μm. Overlaid arrows indicate the calculated self-focusing length for the interaction conditions. (c) Particle-in-cell simulations illustrating density-dependent collapse location for jet $d_{FWHM}$ =700 μm, and peak densities of 0.4 $n_c$ (left) and 0.25 $n_c$ (right). The pulse (overlaid) collapses on the upramp in the first case, while collapsing just past the peak of the jet in the second case.

Simulations were performed with the particle-in-cell code EPOCH [20] for interactions between a 20 mJ 3.9 μm drive pulse and a 700 μm FWHM target with peak densities ranging from 0.1-0.5 $n_{cr}$. Simulations were in agreement with experimental results, producing few MeV electron beams with total charge ~100 pC. Frames from two different simulations are shown in Fig. 4 (c), demonstrating the shift in collapse position for 0.25 $n_{cr}$ and 0.4 $n_{cr}$ respectively.

In summary, we have observed MeV-scale electron acceleration driven by ultrashort mid-IR pulses for the first time and demonstrated the importance of the relativistic self-focusing length in driving electron acceleration. Operating in the mid-infrared enables near and super critical density interactions with a simple gas jet and sets the stage for schemes that may require more tailored near-critical density profiles. Mid-IR drivers will also benefit the standard bubble regime of laser wakefield acceleration, enabling more detailed imaging of larger bubbles and easier synchronization with secondary laser pulses and electron bunches.

**Funding.** Air Force Office of Scientific Research (AFOSR) (FA95501610284, FA95501610121); U.S. Department of Energy (DOE) (DESC0015516); National Science Foundation (NSF) (PHY1619582); U.S. Department of Homeland Security (DHS) (2016DN077ARI104). D. Woodbury acknowledges support from the DOE NNSA SSGF program, (DE-NA0002135).

**Acknowledgment**. We thank George Hine for helpful discussions.